# An atomic frequency comb memory in rare-earth doped thin-film lithium niobate

Subhojit Dutta[1], Yuqi Zhao[1], Uday Saha[1], Demitry Farfurnik[1], Elizabeth A. Goldschmidt[2,3], Edo Waks[1*]

[1]Department of Electrical and Computer Engineering, Institute for Research in Electronics and Applied Physics, and Joint Quantum Institute, University of Maryland, College Park, MD 20742, USA.
[2]Department of Physics, University of Illinois at Urbana-Champaign, Urbana, IL 61801
[3]US Army Research Laboratory, Adelphi, MD 20783
*edowaks@umd.edu

**Atomic frequency combs memories that coherently store optical signals are a key building block for optical quantum computers[1] and quantum networks[2]. Integrating such memories into compact and chip-scale devices is essential for scalable quantum technology, but to date most demonstrations have been in bulk materials or waveguides with large cross-sections[3–5], or using fabrication techniques not easily adaptable to wafer scale processing[6]. We demonstrate compact chip-integrated atomic frequency comb storage in rare earth doped thin-film lithium niobate. Our optical memory exhibits a broad storage bandwidth exceeding 100 MHz, and optical storage time of over 250 ns. The enhanced optical confinement in this device structure enables three orders of magnitude reduction in optical power as compared to large ion-diffused waveguides for the same Rabi frequency[7]. These compact atomic frequency comb memories pave the way towards scalable, highly efficient, electro-optically tunable quantum photonic systems that can store and manipulate light on a compact chip[3].**

Lithium niobate is an ideal photonic platform for developing integrated optical memories based on rare earth ions[8]. It exhibits a wide transparency window, ranging from the ultraviolet to the telecommunications band, and can be integrated with a broad range of rare-earth ion species (e.g., praseodymium, thulium, erbium, etc.) that are optically active over this entire wavelength range[8–11]. Rare-earth-doped ion diffused waveguides patterned in bulk lithium niobate have been previously used to demonstrate optical quantum memory[3,7,12]. However, these devices have a large mode cross section resulting in low optical intensities, necessitating high optical powers for coherent control and storage using atomic frequency combs. Thin-film lithium niobate provides a more compact device structure that is compatible with wafer scale fabrication[13]. Progress in the processing and fabrication of this material achieved extremely low waveguide loss[14] and high-speed electro-optic modulation[15]. Recent efforts to incorporate thulium into thin film lithium niobate have shown bulk-like optical properties which are promising for engineering a photonic memory[8,11,16]. But atomic frequency comb memories with these materials have yet to be realized.

In this letter we report an atomic frequency comb memory in thin-film lithium niobate. We utilize a thin-film lithium niobate waveguide doped with thulium rare-earth ions. We pattern an atomic frequency comb[1,3] on the absorption spectrum of the thulium ion ensemble by spectral hole-burning and achieve optical storage times exceeding 250 ns[1,3]. Through a photon echo measurement, we attain a coherence time of 700 ± 96 ns, matching values of the bulk materials. Due to tight confinement of the waveguide, the ions in the thin film exhibit a three orders of magnitude enhancement in Rabi frequency as compared to ion-diffused waveguides[7]. Such integrated optical memories could serve as fundamental components for high bandwidth quantum information processors on-a-chip[17,18].

Fig. 1a shows a schematic of the device structure, which consists of a ridge waveguide patterned in thin-film lithium niobate. The thin-film lithium niobate is doped with thulium ions density of 0.1%. Fig. 1b shows a scanning electron microscope image of a fabricated waveguide with grating coupler to couple light in and out of the chip. The fabrication details were reported in earlier work[7] and are also provided in supplementary section S1. The fabricated waveguide is 0.8 mm long, corresponding to an optical depth of 1 for the thulium ion absorption. Fig. 1c shows the calculated cross-sectional mode profile of the simulated TE mode in the thin film waveguide, which is aligned along the *x* axis of the crystal which achieves maximal rare-earth ion absorption. The optical mode has a transverse area of (176 nm X 400 nm) 0.07 µm$^2$, (measured as the full width at half maxima.) which is three orders of magnitude smaller than ion-diffused waveguides[7]. This leads to strong light confinement and high field intensities within the waveguide.

We first characterize the coherence time of the atomic ensemble, which puts a fundamental limit on the storage time of the medium. We perform this measurement using a photon echo pulse sequence, shown in Fig. 2a. The pulse sequence uses two laser pulses to coherently rotate the atomic population. The first laser pulse has a pulse area corresponding to a $\pi/2$ pulse, which coherently rotates the population to a superposition of the ground and excited state. The second pulse, delayed by time $\tau$, has twice the pulse duration resulting in a $\pi$-pulse that rephases the atomic population. This rephasing results in a photon echo delayed by a time $\tau$ relative to the rephasing pulse. By measuring the strength of the photon echo as a function of $\tau$ we can determine the coherence time of the rare-earth ensemble.

We generate the rotation pulses by modulating the intensity of a resonant continuous wave laser (see supplementary section S2 for additional details). To determine the pulse intensities that achieve the $\pi$ and $\pi/2$ condition, we plot the power of the output photon echo as a function of rephasing pulse duration, keeping the first pulse duration fixed. Fig. 2b shows the measured results which exhibit Rabi oscillation with an optimal $\pi$ pulse duration of 70 ns. From the pulse width of the $\pi$ pulse, we determine that the Rabi frequency of the dipole transition is 44.9 MHz ($\pi/t_2$) for a driving power of 1 µW in the waveguide (see supplementary section S3 for additional details). This power is three orders of magnitude smaller than the power require in to achieve the same Rabi frequency in ion diffused waveguides[7].

Fig. 2c shows the photon echo amplitude as a function of the delay $\tau$. The echo decays with an exponential time constant of 350 ± 48 ns. From this measurement we determine the coherence time $T_2 = 2\tau$ to be 700 ± 96 ns, which corresponds to a homogeneous linewidth of 450 kHz. This linewidth is comparable to previous measurements for thulium ions in bulk lithium niobate at 4 K[19], indicating that thin-film processing does not significantly degrade the coherence properties of the rare-earth ions.

To realize an atomic frequency comb memory, we implement the pulse sequence illustrated in Fig. 3a composed of a periodic pulse train. The pulse train burns a series of spectral holes at fixed frequency intervals, thereby creating an atomic frequency comb[1,3]. When a pulse enters the prepared atomic ensemble, it excites the individual comb teeth which rapidly dephase, and then rephase at a time $2\pi/\Delta$ and re-emit the photon. We burn the comb using a pulse train of 150 pulses with 10 ns pulse duration and variable period $T = 2\pi/\Delta$ that sets the storage time. We set the peak power of the comb sequence at 0.5 µW in the waveguide (see supplementary section S3 for a detailed discussion).

Fig. 3b shows the absorption spectrum of the atomic ensemble after the comb burning pulse sequence using a burn period of $T = 130$ ns. We determine this absorption spectrum by measuring the transmission of a weak probe pulse, attenuated by three orders of magnitude relative to the burn pulse. We sweep the frequency of the probe and measure the light at the output grating. The absorption spectrum features a series of absorption lines corresponding to the different comb teeth. The combs spacing is 6.3 MHz, as expected for the selected pulse periodicity (see supplementary section S4 for a detailed discussion). The measured atomic frequency comb has an optical depth contrast of 0.23. For line 1 indicated in the figure, the full-width half-maximum linewidth is 3.7 MHz, corresponding to a comb finesse of 1.7. The full-width half-maximum linewidth and the corresponding finesse of line 2 labeled in the figure is 3.6 MHz and 1.75 respectively. The rise in the intensity over the different comb teeth is due to the spectrum of the input pulse, which is induced by the acousto-optic modulator bandwidth of 50 MHz.

To store light in the atomic frequency comb memory, we inject a 10 ns optical pulse at 794.2 nm, the center frequency of the atomic frequency comb. Fig. 3c shows that waveguide output for several different frequency combs corresponding to storage times ranging from 90 ns to 250 ns. Each comb results in an output pulse that is delayed by the correct time programmed into the frequency comb memory. We calculate the storage efficiency of the comb, defined as the ratio of the input and output pulse energies (integrated counts), to be 0.14% for a 90 ns storage time. This storage is primarily limited by the small optical depth contrast that we can achieve. The storage efficiency decreases as we increase the storage time of the system. For a storage time of 130 ns the efficiency decreases to 0.09%, while at 250 ns it further reduces to 0.03%. We note that the storage efficiency does not exhibit a single exponential decay, as would be expected if pure dephasing process of the thulium transition was dominating the dependence on storage time. We attribute this non-exponential decay to the fact that the finesse of the comb teeth varies significantly with increasing storage time due to the reduced free spectral range bringing the comb teeth closer.

In conclusion, we demonstrated an atomic frequency comb memory in thulium-doped thin film lithium niobate. The tight confinement of the thin-film lithium niobate waveguide led to a three orders of magnitude improvement in the Rabi frequency and we observed no degradation of the coherence properties of the thulium dopants compared to bulk materials. We can further improve the storage efficiency by burning the frequency comb in the presence of a magnetic field and using the Zeeman split levels to pattern the absorption spectrum, thereby increasing the optical depth[20] or by using a longer waveguide. As a further extension, we can take advantage of the thin film platform and design impedance matched cavities[14,21,22] as a pathway to achieving a unit efficiency optical memory on chip[23]. The ability to integrate active modulation on chip could further enable electro-optically tunable[24,25] integrated photonic memory. Such a versatile platform is therefore a critical step towards scalable, highly efficient, electro-optically tunable quantum photonic systems where one can store and manipulate light on chip with high bandwidth and low powers.

SD, EW and EAG conceived the experiment. SD and US fabricated the device. SD and YZ performed the experiment. DF supported in setting up the experiment. SD and YZ analyzed the data. SD, EW, EAG and YZ wrote the manuscript.

The authors declare no competing interests.

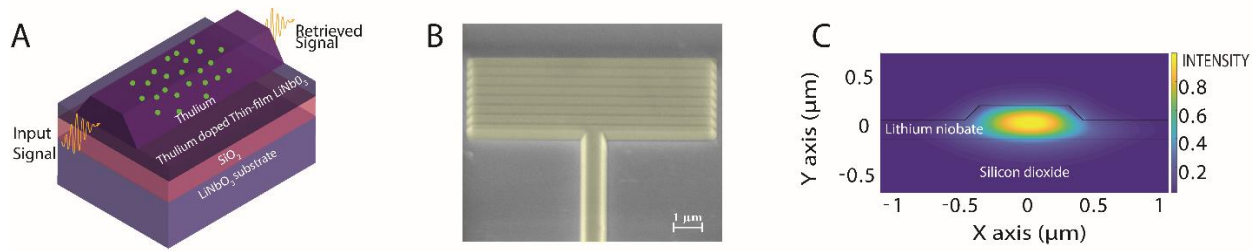

Figure 1. (A) Schematic of the device structure, composed of a thin-film lithium niobate waveguide doped with thulium rare-earth ions. (B) Scanning electron microscopy image of the waveguide with grating couplers. (C) Finite difference time domain simulation showing the $x$ component of the electric field of the waveguide mode, which is maximally aligned with the rare-earth ion absorption.

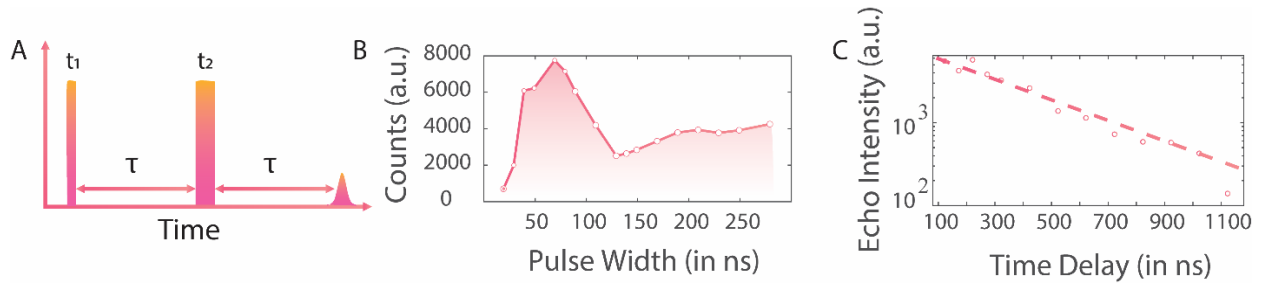

Figure 2. (A) The pulse sequence for performing photon echo measurement. The two pulses have duration of $t_1$ and $t_2$ respectively, which are selected to create a $\pi/2$ and $\pi$ pulse rotation on the atomic ensemble. (B) Intensity of the photon echo as a function of $t_2$. The oscillation corresponds to coherent rotation of the atomic ensemble, with the $\pi$ pulse condition achieved at a pulse duration of 70 ns. (C) Photon echo intensity as a function of delay $(\tau + \frac{t1+t2}{2})$ (red dots), along with an exponential fit (dashed line). The exponential fit decays with a time constant of $350 \pm 48$ ns, which corresponds to a coherent time of $700 \pm 96$ ns.

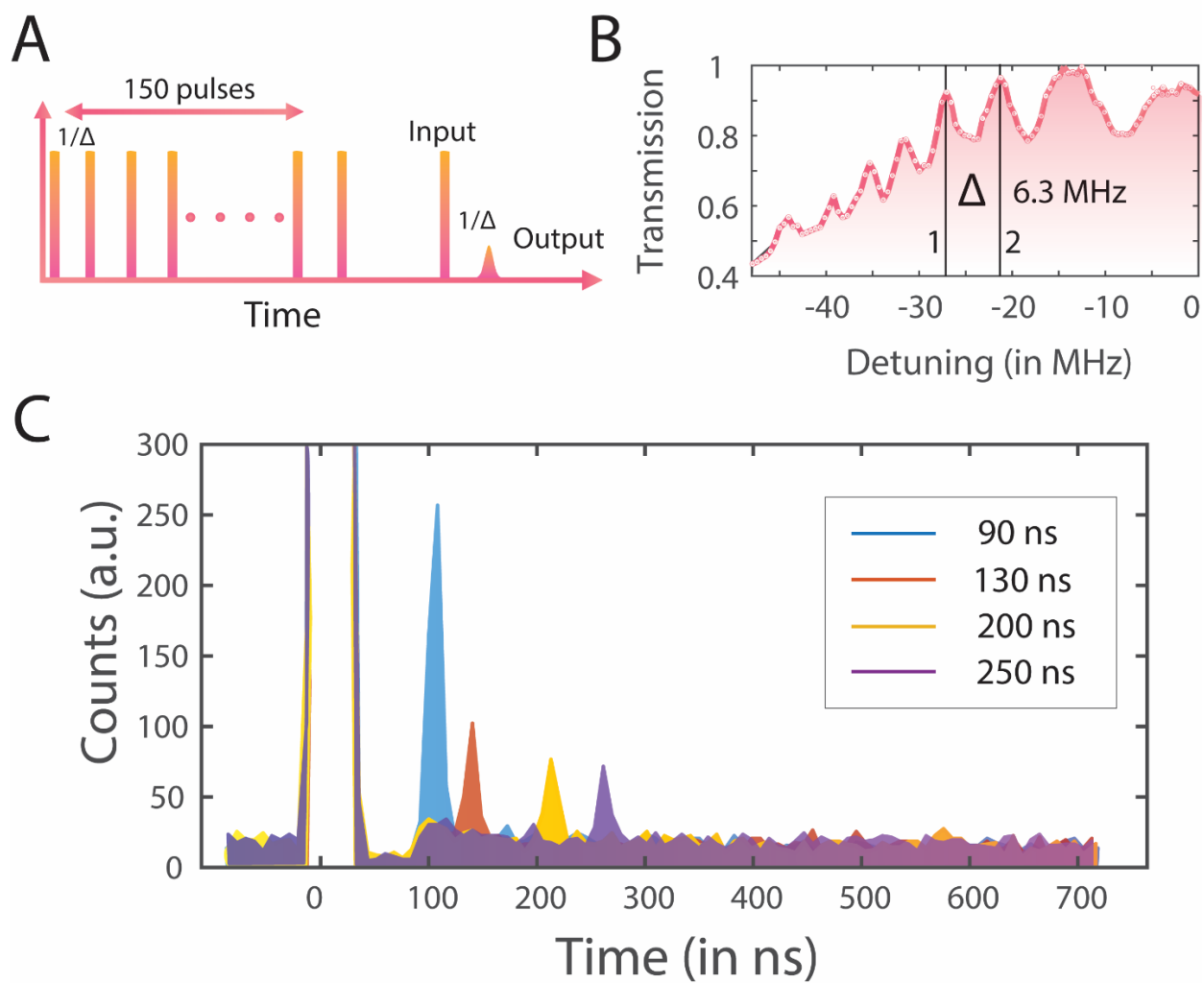

Figure 3. (A) Pulse sequence used to generate an atomic frequency comb. Each pulse is 10 ns in duration. (B) Transmission intensity of a tunable probe pulse through the atomic frequency comb as a function of probe frequency. The oscillations in the transmission correspond to the comb teeth burned by the pulse sequence illustrated in panel A. (C) Demonstration of photon storage for various periodicities of the burn pulse sequence. In each case, the comb releases the probe pulse at a time delay given by the periodicity of the frequency comb.

# Supplementary Information

**S1. Fabrication**

We begin with a bulk X-cut lithium niobate substrate doped with 0.1% $Tm^{3+}$. Using a commercial smart-cut process (NANOLN), we fabricate a 300 nm layer of doped single crystal lithium niobate wafer bonded to a 2 µm thick layer of silicon dioxide grown on undoped bulk lithium niobate as the substrate material. We etch waveguides into the doped thin film using a home developed two-layer electron beam lithography and dry etching process.

## S2. Experimental Setup

To perform the measurements, we use two fiber coupled acousto-optic modulators (Brimrose Corp.) to carve out pulses at two independent frequencies, $f_0$ and $f_1$, from the single frequency input laser (M2 Solstis). We can switch the modulators by means of an RF switch (Mini circuits) and control pulses generated by an arbitrary waveform generator. We drive the AOMs using a pair of Voltage Controlled Oscillators which can be tuned to independently control the output frequencies $f_0$ and $f_1$.

## S3. Calibrating Power in the Waveguide

In order to calibrate the power in the waveguide, we measure the end-to-end power transmission through the setup and the device under test. We measure the input power right before the objective lens we use for focusing on the sample grating couplers and measure the output power scattered from the output grating into a single mode fiber. We assume that the in-coupling and out-coupling efficiencies are equal, and we have negligible waveguide loss for sub mm long waveguides. Using such a procedure we extract the input and output coupling efficiency as 0.1% each. In this way we have a good order of magnitude estimate of the power coupled into the waveguide.

## S4. Atomic Frequency Comb Preparation

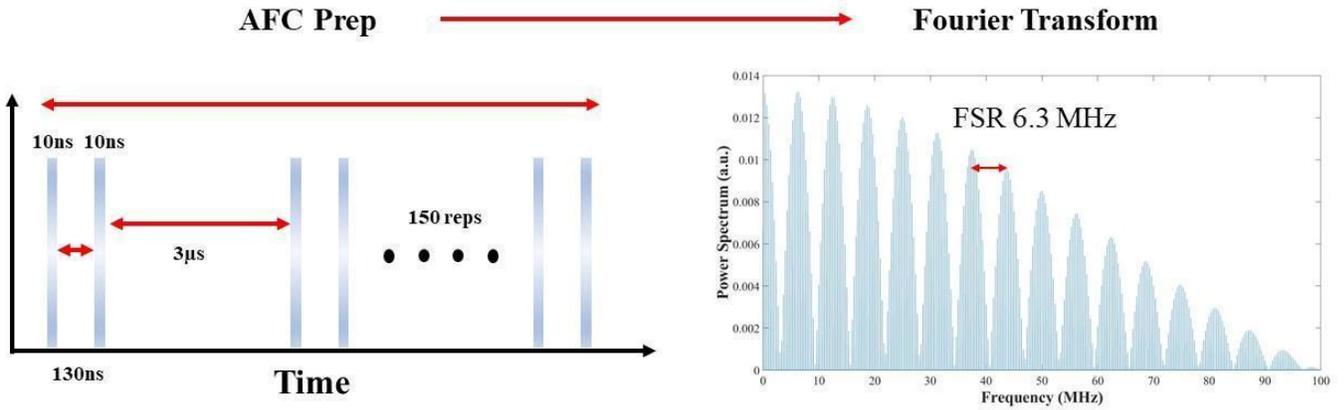

Figure S3. The schematic on the left shows a time domain representation of the pulse sequence for preparing an atomic frequency comb. The figure on the right demonstrates a Fourier transform of the same pulse sequence representing a frequency comb in the Fourier space.

To prepare the atomic frequency comb, we send a series of 10 ns long pulse pairs separated by the programmable storage time (130 ns in Fig S3). The wait time between the pulse pairs is 3 μs, which is long compared to coherence time of the ions. We use a series of 150 repetitions to burn a series of spectral holes. The Fourier transform of this pulse sequence consist of a series of frequency peaks separated by 6.3 MHz. This results in a spectral hole every 6.3 MHz which matches well with the experimentally observed free spectral range.